# Tuffy: Scaling up Statistical Inference in Markov Logic Networks using an RDBMS*

Feng Niu      Christopher Ré      AnHai Doan      Jude Shavlik

University of Wisconsin-Madison
{leonn,chrisre,anhai,shavlik}@cs.wisc.edu

## ABSTRACT

Markov Logic Networks (MLNs) have emerged as a powerful framework that combines statistical and logical reasoning; they have been applied to many data intensive problems including information extraction, entity resolution, and text mining. Current implementations of MLNs do not scale to large real-world data sets, which is preventing their widespread adoption. We present Tuffy that achieves scalability via three novel contributions: (1) a bottom-up approach to grounding that allows us to leverage the full power of the relational optimizer, (2) a novel hybrid architecture that allows us to perform AI-style local search efficiently using an RDBMS, and (3) a theoretical insight that shows when one can (exponentially) improve the efficiency of stochastic local search. We leverage (3) to build novel partitioning, loading, and parallel algorithms. We show that our approach outperforms state-of-the-art implementations in both quality and speed on several publicly available datasets.

## 1. INTRODUCTION

Over the past few years, Markov Logic Networks (MLNs) have emerged as a powerful and popular framework combining logical and probabilistic reasoning. MLNs have been successfully applied to a wide variety of data management problems, e.g., information extraction, entity resolution, and text mining. In contrast to probability models like factor graphs [23] that require complex distributions to be specified in tedious detail, MLNs allow us to declare a rigorous statistical model at a much higher conceptual level using first-order logic. For example, to classify papers by research area, one could write a rule such as *"it is likely that if one paper cites another they are in the same research area."*

*We gratefully acknowledge the support of Defense Advanced Research Projects Agency (DARPA) Machine Reading Program under Air Force Research Laboratory (AFRL) prime contract no. FA8750-09-C-0181. Any opinions, findings, and conclusion or recommendations expressed in this material are those of the authors and do not necessarily reflect the view of DARPA, AFRL, or the US government.

Our interest in MLNs stems from our involvement in a DARPA project called "Machine Reading." The grand challenge is to build software that can read the Web, i.e., extract and integrate structured data (e.g., entities, relationships) from Web data, then use this structured data to answer user queries. The current approach is to use MLNs as a lingua franca to combine many different kinds of extractions into one coherent picture. To accomplish this goal, it is critical that MLNs scale to large data sets.

Unfortunately, none of the current MLN implementations scale beyond relatively small data sets (and even on many of these data sets, existing implementations routinely take hours to run). The first obvious reason is that these are *in-memory* implementations: when manipulating large intermediate data structures that overflow main memory, they either crash or thrash badly. Consequently, there is an emerging effort across several research groups to scale up MLNs. In this paper, we describe our system, Tuffy[1], that leverages an RDBMS to address the above scalability and performance problems.

There are two aspects of MLNs: learning and inference [21]. We focus on inference, since typically a model is learned once, and then an application may perform inference many times using the same model; hence inference is an on-line process, which must be fast. Moreover, MLN learning algorithms typically invoke inference as a subroutine repeatedly. Conceptually, inference[2] in MLNs has two phases: a *grounding phase*, which constructs a large, weighted SAT formula, and a *search phase*, which searches for a low cost (weight) assignment (called a *solution*) to the SAT formula from grounding (using WalkSAT [13], a local search procedure). Grounding is a non-trivial portion of the overall inference effort: on a classification benchmark (called RC) the state-of-the-art MLN inference engine, Alchemy [7], spends over 96% of its execution time in grounding. The state-of-the-art strategy for the grounding phase (and the one used by Alchemy) is a top-down procedure (similar to the proof strategy in Prolog). In contrast, we propose a bottom-up grounding strategy. Intuitively, bottom-up grounding allows Tuffy to fully exploit the RDBMS optimizer, and thereby significantly speed up the grounding phase of MLN inference. On an entity resolution task, Alchemy takes over 7 hours to complete grounding, while Tuffy's grounding finishes in less than 2 minutes.

---

[1] http://www.cs.wisc.edu/hazy/tuffy/
[2] We focus on *maximum a posteriori inference* which is critical for many integration tasks. We discuss marginal inference in Appendix A.5.

But not all phases are well-optimized by the RDBMS: during the search phase, we found that the RDBMS implementation performed poorly. The underlying reason is a fundamental problem for pushing local search procedures into an RDBMS: search procedures often perform *inherently sequential, random* data accesses. Consequently, any RDBMS-based solution must execute a large number of disk accesses, each of which has a substantial overhead (due to the RDBMS) versus direct main-memory access. Not surprisingly, given the same amount of time, an in-memory solution can execute between three and five orders of magnitude more search steps than an approach that uses an RDBMS. Thus, to achieve competitive performance, we developed a novel hybrid architecture that supports local search procedures in main memory whenever possible. This is our second technical contribution.

Our third contribution is a simple partitioning technique that allows TUFFY to introduce parallelism and use less memory than state-of-the-art approaches. Surprisingly, this same technique often allows TUFFY to speed up the search phase exponentially. The underlying idea is simple: in many cases, a local search problem can be divided into multiple independent subproblems. For example, the formula that is output by the grounding phase may consist of multiple connected components. On such datasets, we derive a sufficient condition under which solving the subproblems independently results in exponentially faster search than running the larger global problem (Thm. 3.1). An application of our theorem shows that on an information extraction testbed, a system that is not aware of this phenomenon (such as ALCHEMY) must take at least $2^{200}$ more steps than TUFFY to reach a solution with the same quality. Empirically we found that, on some real-world datasets, solutions found by TUFFY within one minute have higher quality than those found by non-partitioning systems (such as ALCHEMY) even after running for days.

The exponential difference in running time for independent subproblems versus the larger global problem suggests that in some cases, further decomposing the search space may improve the overall runtime. To implement this idea for MLNs, we must address two difficult problems: (1) partitioning the formula from grounding (and so the search space) to minimize the number of formula that are split between partitions, and (2) augmenting the search algorithm to be aware of partitioning. We show that the first problem is NP-hard (even to approximate), and design a scalable heuristic partitioning algorithm. For the second problem, we apply a technique from non-linear optimization to leverage the insights gained from our characterization of the phenomenon described above. The effect of such partitioning is dramatic. As an example, on a classification benchmark (called RC), TUFFY (using 15MB of RAM) produces much better result quality in minutes than ALCHEMY (using 2.8GB of RAM) even after days of running. In fact, TUFFY is able to answer queries on a version of the RC dataset that is over two orders of magnitude larger. (We estimate that ALCHEMY would need 280GB+ of RAM to process it.)

*Related Work.* MLNs are an integral part of state-of-the-art approaches in a variety of applications: natural language processing [22], ontology matching [29], information extraction [18], entity resolution [25], etc. And so, there is an application push to support MLNs.

Pushing statistical reasoning models inside a database system has been a goal of many projects [5, 10, 11, 20, 27]. Most closely related is the BAYESSTORE project, in which the database essentially stores *Bayes Nets* [17] and allows these networks to be retrieved for inference by an external program. In contrast, TUFFY uses an RDBMS to optimize the inference procedure. The Monte-Carlo database [10] made sampling a first-class citizen inside an RDBMS. In contrast, in TUFFY our approach can be viewed as pushing classical search inside the database engine. One way to view an MLN is a compact specification of *factor graphs* [23]. Sen et al. [23] proposed new algorithms; in contrast, we take an existing, widely used class of algorithms (local search), and our focus is to leverage the RDBMS to improve performance.

There has also been an extensive amount of work on *probabilistic databases* [1, 2, 4, 19] that deal with simpler probabilistic models. Finding the most likely world is trivial in these models; in contrast, it is highly non-trivial in MLNs (in fact, it is NP-hard [6]). Finally, none of these prior approaches deal with the core technical challenge TUFFY addresses, which is handling AI-style search inside a database. Additional related work can be found in Appendix D.

*Contributions, Validation, and Outline.* To summarize, we make the following contributions:

- In Section 3.1, we design a solution that pushes MLNs into RDBMSes. The key idea is to use bottom-up grounding that allows us to leverage the RDBMS optimizer; this idea improves the performance of the grounding phase by several orders of magnitude.

- In Section 3.2, we devise a novel hybrid architecture to support efficient grounding and in-memory inference. By itself, this architecture is far more scalable and, given the same amount of time, can perform orders of magnitude more search steps than prior art.

- In Section 3.3, we describe novel data partitioning techniques to decrease the memory usage and to increase parallelism (and so improve the scalability) of TUFFY's in-memory inference algorithms. Additionally, we show that for any MLN with an MRF that contains multiple components, partitioning could exponentially improve the expected (average case) search time.

- In Section 3.4, we generalize our partitioning results to arbitrary MLNs using our characterization of the partitioning phenomenon. These techniques result in our highest quality, most space-efficient solutions.

We present an extensive experimental study on a diverse set of MLN testbeds to demonstrate that our system TUFFY is able to get better result quality more quickly and work over larger datasets than the state-of-the-art approaches.

## 2. PRELIMINARIES

We illustrate a Markov Logic Network program using the example of classifying papers by topic area. We then define the semantics of MLNs and the mechanics of inference.

### 2.1 The Syntax of MLNs

Figure 1 shows an example input MLN program for TUFFY that is used to classify paper references by topic area, such as databases, systems, AI, etc. In this example, a user gives

| | weight | rule | | |
|---|---|---|---|---|
| paper(PaperID, URL) | 5 | $cat(p, c1), cat(p, c2) => c1 = c2$ | ($F_1$) | wrote('Joe', 'P1') |
| wrote(Author, Paper) | 1 | $wrote(x, p1), wrote(x, p2), cat(p1, c) => cat(p2, c)$ | ($F_2$) | wrote('Joe', 'P2') |
| refers(Paper, Paper) | 2 | $cat(p1, c), refers(p1, p2) => cat(p2, c)$ | ($F_3$) | wrote('Jake', 'P3') |
| cat(Paper, Category) | $+\infty$ | $paper(p, u) => \exists x. wrote(x, p)$ | ($F_4$) | refers('P1', 'P3') |
| | -1 | $cat(p, \text{'Networking'})$ | ($F_5$) | cat('P2', 'DB') |
| Schema | | A Markov Logic Program | | Evidence |

Figure 1: A Sample Markov Logic Program: The goal is to classify papers by area. As evidence we are given author and citation information of all papers, as well as the labels of a subset of the papers; we want to classify the remaining papers. Any variable not explicitly quantified is universally quantified.

TUFFY a set of relations that capture information about the papers in her dataset: she has extracted authors and citations and stored them in the relations wrote(Author,Paper) and refers(Paper,Paper). She may also provide *evidence*, which is data that she knows to be true (or false). Here, the evidence shows that Joe wrote papers P1 and P2 and P1 cited another paper P3. In the relation cat (for 'category'), she provides TUFFY with a subset of papers and the categories into which they fall. The cat relation is incomplete: some papers are not labeled. We can think of each possible labeling of these papers as an instantiation of the cat relation, which can be viewed as a *possible world* [8]. The classification task is to find the most likely labeling of papers by topic area, and hence the most likely possible world.

To tell the system which possible world it should produce, the user provides (in addition to the above data) a set of rules that incorporate her knowledge of the problem. A simple example rule is $F_1$:

$$cat(p, c1), cat(p, c2) => c1 = c2 \quad (F_1)$$

Intuitively, $F_1$ says that a paper should be in one category. In MLNs, this rule may be *hard*, meaning that it behaves like a standard key constraint: in any possible world, each paper must be in at most one category. This rule may also be *soft*, meaning that it may be violated in some possible worlds. For example, in some worlds a paper may be in two categories. Soft rules also have *weights* that intuitively tell us how likely the rule is to hold in a possible world. In this example, $F_1$ is a soft rule and has weight 5. Roughly, this means that a fixed paper is at least $e^5$ times more likely to be in a single category compared to being in multiple categories. MLNs can also involve data in non-trivial ways, we refer the reader to Appendix A.1 for a more complete exposition.

*Query Model.* Given the data and the rules, a user may write arbitrary queries in terms of the relations. In TUFFY, the system is responsible for filling in whatever missing data is needed: in this example, the category of each unlabeled paper is unknown, and so to answer a query the system infers the most likely labels for each paper from the evidence.

## 2.2 Semantics of MLNs

We describe the semantics of MLNs. Formally, we first fix a schema $\sigma$ (as in Figure 1) and a domain $D$. Given as input a set of formula $\bar{F} = F_1, \ldots, F_N$ (in *clausal form*[3]) with weights $w_1, \ldots, w_N$, they define a probability distribution over *possible worlds* (deterministic databases). To construct this probability distribution, the first step is *grounding*: given a formula $F$ with free variables $\bar{x} = (x_1, \cdots, x_m)$, then for each $\bar{d} \in D^m$, we create a new formula $g_{\bar{d}}$ called a *ground clause* where $g_{\bar{d}}$ denotes the result of substituting each variable $x_i$ of $F$ with $d_i$. For example, for $F_3$ the variables are $\{p1, p2, c\}$: one tuple of constants is $\bar{d} = $ ('P1','P2','DB') and the ground formula $f_{\bar{d}}$ is:

$$cat(\text{'P1'}, \text{'DB'}), refers(\text{'P1'}, \text{'P2'}) => cat(\text{'P2'}, \text{'DB'})$$

Each constituent in the ground formula, such as cat('P1', 'DB') and refers('P1', 'P2'), is called a *ground predicate* or *atom* for short. In the worst case there are $D^3$ ground clauses for $F_3$. For each formula $F_i$ (for $i = 1 \ldots N$), we perform the above process. Each ground clause $g$ of a formula $F_i$ is assigned the same weight, $w_i$. So, a ground clause of $F_1$ has weight 5, while any ground clause of $F_2$ has weight 1. We denote by $G = (\bar{g}, w)$ the set of all ground clauses of $\bar{F}$ and a function $w$ that maps each ground clause to its assigned weight. Fix an MLN $\bar{F}$, then for any possible world (instance) $I$ we say a ground clause $g$ is *violated* if $w(g) > 0$ and $g$ is false in $I$ or if $w(g) < 0$ and $g$ is true in $I$. We denote the set of ground clauses violated in a world $I$ as $V(I)$. The cost of the world $I$ is

$$\text{cost}(I) = \sum_{g \in V(I)} |w(g)| \quad (1)$$

Through cost, an MLN defines a probability distribution over all instances (denoted Inst) as:

$$\Pr[I] = Z^{-1} \exp\{-\text{cost}(I)\} \text{ where } Z = \sum_{J \in \text{Inst}} \exp\{-\text{cost}(J)\}$$

A lowest cost world $I$ is called a *most likely world*. Since $\text{cost}(I) \geq 0$, if $\text{cost}(I) = 0$ then $I$ is a most likely world. On the other hand the most likely world may have positive cost. There are two main types of inference with MLNs: MAP (maximum a posteriori) inference, where we want to find a most likely world, and marginal inference, where we want to compute marginal probabilities. TUFFY is capable of both types of inference, but we present only MAP inference in the body of this paper. We refer the reader to Appendix A.5 for details of marginal inference.

## 2.3 Inference

We now describe the state of the art of inference for MLNs (as in ALCHEMY, the reference MLN implementation).

*Grounding.* Conceptually, to obtain the ground clauses of an MLN formula $F$, the most straightforward way is to enumerate all possible assignments to the free variables in $F$. There have been several heuristics in the literature that improve the grounding process by pruning groundings that have no effect on inference results; we describe the heuristics that TUFFY (and ALCHEMY) implements in Appendix A.3.

---
[3]Clausal form is a disjunction of positive or negative literals. For example, the rule is $R(a) => R(b)$ is not in clausal form, but is equivalent to $\neg R(a) \vee R(b)$, which is in clausal form.

The set of ground clauses corresponds to a hypergraph where each atom is a node and each clause is a hyperedge. This graph structure is often called a *Markov Random Field* (MRF). We describe this structure formally in Appendix A.2.

*Search.* Finding a most likely world of an MLN is a generalization of the (NP-hard) MaxSAT problem. In this paper we concentrate on one of the most popular heuristic search algorithms, WalkSAT [13], which is used by ALCHEMY. WalkSAT works by repeatedly selecting a random violated clause and "fixing" it by flipping (i.e., changing the truth value of) an atom in it (see Appendix A.4). As with any heuristic search, we cannot be sure that we have achieved the optimal, and so the goal of any system that executes such a search procedure is: *execute more search steps in the same amount of time.*

*Problem Description.* The primary challenge that we address in this paper is scaling both phases of MAP inference algorithms, grounding and search, using an RDBMS. Second, our goal is to improve the number of (effective) steps of the local search procedure using parallelism and partitioning – but only when it provably improves the search quality. To achieve these goals, we attack three main technical challenges: (1) efficiently grounding large MLNs, (2) efficiently performing inference (search) on large MLNs, and (3) designing partitioning and partition-aware search algorithms that preserve (or enhance) search quality and speed.

## 3. TUFFY SYSTEMS

In this section, we describe our technical contributions: a bottom-up grounding approach to fully leverage the RDBMS (Section 3.1); a hybrid main-memory RDBMS architecture to support efficient end-to-end inference (Section 3.2); and data partitioning which dramatically improves TUFFY's space and time efficiency (Section 3.3 and Section 3.4).

### 3.1 Grounding with a Bottom-up Approach

We describe how TUFFY performs grounding. In contrast to top-down approaches (similar to Prolog) that employ nested loops and that is used by prior MLN systems such as ALCHEMY, Tuffy takes a bottom-up approach (similar to Datalog) by expressing grounding as a sequence of SQL queries. Each SQL query is optimized by the RDBMS, which allows TUFFY to complete the grounding process orders of magnitude more quickly than prior approaches.

For each predicate $P(\bar{A})$ in the input MLN, TUFFY creates a relation $R_P(\underline{aid}, \bar{A}, truth)$ where each row $a_p$ represents an atom, $aid$ is a globally unique identifier, $\bar{A}$ is the tuple of arguments of $P$, and $truth$ is a three-valued attribute that indicates if $a_p$ is true or false (in the evidence), or not specified in the evidence. These tables form the input to grounding, and TUFFY constructs them using standard bulk-loading techniques.

In TUFFY, we produce an output table $C(\underline{cid}, lits, weight)$ where each row corresponds to a single ground clause. Here, $cid$ is the id of a ground clause, $lits$ is an array that stores the atom id of each literal in this clause (and whether or not it is negated), and $weight$ is the weight of this clause. We first consider a formula without existential quantifiers. In this case, the formula $F$ can be written as $F(\bar{x}) = l_1 \vee \cdots \vee l_N$ where $\bar{x}$ are all variables in $F$. TUFFY produces a SQL query $Q$ for $F$ that joins together the relations corresponding to the predicates in $F$ to produce the atom ids of the ground clauses (and whether or not they are negated). The join conditions in $Q$ enforce variable equality inside $F$, and incorporate the pruning strategies described in Appendix A.3. For more details on the compilation procedure see Appendix B.1.

### 3.2 A Hybrid Architecture for Inference

Our initial prototype of Tuffy runs both grounding and search in the RDBMS. While the grounding phase described in the previous section has good performance and scalability, we found that performing search in an RDBMS is often a bottleneck. Thus, we design a hybrid architecture that allows efficient in-memory search (in Java) while retaining the performance benefits of RDBMS-based grounding. To see why in-memory search is critical, recall that WalkSAT works by selecting an unsatisfied clause $C$, selecting an atom in $C$, and "flipping" that atom to satisfy $C$. Thus, Walk-SAT performs a large number of random accesses to the data representing ground clauses and atoms. Moreover, the data that is accessed in one iteration depends on the data that is accessed in the previous iteration. And so, this access pattern prevents both effective caching and parallelism, which causes a high overhead per data access. Thus, we implement a hybrid architecture where the RDBMS performs grounding and TUFFY is able to read the result of grounding from the RDBMS into memory and perform inference. If the grounding result is too large to fit in memory, TUFFY invokes an implementation of search directly inside the RDBMS (Appendix B.2). This approach is much less efficient than in-memory search, but it runs on datasets larger than main memory without crashing. Appendix B.3 illustrates the architecture of TUFFY in more detail.

While it is clear that this hybrid approach is at least as scalable as a direct memory implementation (such as ALCHEMY), there are in fact cases where TUFFY can run in-memory search whereas ALCHEMY would crash. The reason is that the space requirement of a purely in-memory implementation is determined by the peak memory footprint throughout grounding *and* search, whereas TUFFY needs main memory only for search. For example, on a dataset called Relational Classification (RC), ALCHEMY allocated 2.8 GB of RAM only to produce 4.8 MB of ground clauses. On RC, TUFFY uses only 19 MB of RAM.

### 3.3 Partitioning to Improve Performance

In the following two sections, we study how to further improve TUFFY's space and time efficiency without sacrificing its scalability. The underlying idea is simple: we will try to partition the data. By splitting the problem into smaller pieces, we can reduce the memory footprint and introduce parallelism, which conceptually breaks the sequential nature of the search. These are expected benefits of partitioning. An unexpected benefit is an exponentially increase of the effective search speed, a point that we return to below.

First, observe that the logical forms of MLNs often result in an MRF with multiple disjoint components (see Appendix B.4). For example, on the RC dataset there are 489 components. Let $G$ be an MRF with components $G_1, \cdots, G_k$; let $I$ be a truth assignment to the atoms in $G$ and $I_i$ its pro-jection over $G_i$. Then, it's clear that $\forall I$

$$\text{cost}^G(I) = \sum_{1 \leq i \leq k} \text{cost}^{G_i}(I_i).$$

Hence, instead of minimizing $\text{cost}^G(I)$ directly, it suffices to minimize each individual $\text{cost}^{G_i}(I_i)$. The benefit is that, even if $G$ itself does not fit in memory, it is possible that each $G_i$ does. As such, we can solve each $G_i$ with in-memory search one by one, and finally merge the results together. [4]

Component detection is done after the grounding phase and before the search phase, as follows. We maintain an in-memory union-find structure over the nodes, and scan the clause table while updating this union-find structure. The result is the set of connected components in the MRF. An immediate issue raised by partitioning is I/O efficiency.

*Efficient Data Loading.* Once an MRF is split into components, loading in and running inference on each component sequentially one by one may incur many I/O operations, as there may be many partitions. For example, the MRF of the Information Extraction (IE) dataset contains thousands of 2-cliques and 3-cliques. One solution is to group the components into batches. The goal is to minimize the total number of batches (and thereby the I/O cost of loading), and the constraint is that each batch cannot exceed the memory budget. This is essentially the bin packing problem, and we implement the *First Fit Decreasing* algorithm [26]. Once the partitions are in memory, we can take advantage of parallelism. We use a round-robin scheduling policy.

*Improving Search Speed using Partitioning.* Although processing each component individually produces solutions that are no worse than processing the whole graph at once, we give an example to illustrate that component-aware processing may result in exponentially faster speed of search.

**Example 1** Consider an MRF consisting of $N$ identical connected components each containing two atoms $\{X_i, Y_i\}$ and three weighted clauses

$$\{(X_i, 1), (Y_i, 1), (X_i \vee Y_i, -1)\},$$

where $i = 1 \ldots N$ and the second component of each tuple is the weight. Based on how WalkSAT works, it's not hard to show that, if $N = 1$, starting from a random state, the expected hitting time[5] of the optimal state, i.e., $X_1 = Y_1 = True$, is no more than 4. Therefore, if we run WalkSAT on each component separately, the expected runtime of reaching the optimum is no more than $4N$. Now consider the case where we run WalkSAT on the whole MRF. Intuitively, reaching the optimal state requires "fixing" suboptimal components one by one. As the number of optimal components increases, however, it becomes more and more likely that one step of WalkSAT "breaks" an optimal component instead of fixing a suboptimal component. Such check and balance makes it very difficult for WalkSAT to reach the optimum. Indeed, Appendix B.5 shows that the expected hitting time is at least $2^N$ – an exponential gap!

---

[4] ALCHEMY exploits knowledge-based model construction (KBMC) [28] to find the minimal subgraph of the MRF that is needed for a given query. ALCHEMY, however, does not use the fact that the MRF output by KBMC may contain several components.

[5] The hitting time is a standard notion from Markov Chains [9], it is a random variable for the number of steps taken by WalkSAT to reach an optimum for the first time.

Let $G$ be an MRF with components $G_1, \ldots, G_N$. *Component-aware WalkSAT* runs WalkSAT except that for each $G_i$, it keeps track of the lowest-cost state it has found so far on that $G_i$. In contrast, regular WalkSAT simply keeps the best overall solution it has seen so far. For $i = 1, \ldots, N$, let $O_i$ be the set of optimal states of $G_i$, and $S_i$ the set of non-optimal states of $G_i$ that differ only by one bit from some $x^* \in O_i$; let $P_i(x \to y)$ be the transition probability of WalkSAT running on $G_i$, i.e., the probability that one step of WalkSAT would take $G_i$ from $x$ to $y$. Given $x$, a state of $G_i$, denote by $v_i(x)$ the number of violated clauses in $G_i$ at state $x$; define

$$\alpha_i(x) = \sum_{y \in O_i} P_i(x \to y), \quad \beta_i(x) = \sum_{y \in S_i} P_i(x \to y).$$

For any non-empty subset $H \subseteq \{1, \ldots, N\}$, define

$$r(H) = \frac{\min_{i \in H} \min_{x \in O_i} v_i(x) \beta_i(x)}{\max_{i \in H} \max_{x \in S_i} v_i(x) \alpha_i(x)}.$$

THEOREM 3.1. *Let $H$ be an arbitrary non-empty subset of $\{1, \ldots, N\}$ s.t. $|H| \geq 2$ and $r = r(H) > 0$. Then, in expectation, WalkSAT on $G$ takes at least $2^{|H|r/(2+r)}$ more steps to find an optimal solution than component-aware WalkSAT.*

The proof is in Appendix B.5. In the worst case, there is only one component, or $r(H) = 0$ for every subset of components $H$ (which happens only if there is a zero-cost solution), and partitioning would become pure overhead (but negligible in our experiments). On an information extraction (IE) benchmark dataset, there is some $H$ with $|H| = 1196$ and $r(H) = 0.5$. Thus, the gap on this dataset is at least $2^{200} \approx 10^{60}$. This explains why TUFFY produces lower cost solutions in minutes than non-partition aware approaches such as ALCHEMY produce even after days.

### 3.4 Further Partitioning MRFs

Although our algorithms are more scalable than prior approaches, if the largest component does not fit in memory then we are forced to run the in-RDBMS version of inference, which is inefficient. Intuitively, if the graph is only weakly connected, then we should still be able to get the exponential speed up of partitioning. Consider the following example.

**Example 2** Consider an MRF consisting of two equally sized subgraphs $G_1$ and $G_2$, plus an edge $e = (a, b)$ between them (Figure 2). Suppose that the expected hitting time of Walk-SAT on $G_i$ is $H_i$. Since $H_1$ and $H_2$ are essentially independent, the hitting time of WalkSAT on $G$ could be roughly $H_1 H_2$. On the other hand, consider the following scheme: enumerate all possible truth assignments to one of the boundary variables $\{a, b\}$, say $a$ – of which there are two – and conditioning on each assignment, run WalkSAT on $G_1$ and $G_2$ independently. Clearly, the overall hitting time is no more than $2(H_1 + H_2)$, which is a huge improvement over $H_1 H_2$ since $H_i$ is usually a high-order polynomial or even exponential in the size of $G_i$.

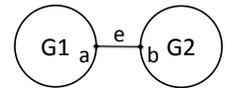

**Figure 2: Ex. 2**

To capitalize on this idea, we need to address two challenges: 1) designing an efficient MRF partitioning algorithm; and 2) designing an effective partition-aware search algorithm. We address each of them in turn.

*MRF Partitioning.* Intuitively, to maximally utilize the memory budget, we want to partition the MRF into roughly equal sizes; to minimize information loss, we want to minimize total weight of clauses that span over multiple partitions, i.e., the *cut size*. To capture this notion, we define a balanced bisection of a hypergraph $G = (V, E)$ as a partition of $V = V_1 \cup V_2$ such that $|V_1| = |V_2|$. The cost of a bisection $(V_1, V_2)$ is $|\{e \in E | e \cap V_1 \neq \emptyset \text{ and } e \cap V_2 \neq \emptyset\}|$.

THEOREM 3.2. *Consider the MLN $\Gamma$ given by the single rule $p(x), r(x,y) \rightarrow p(y)$ where $r$ is an evidence predicate. Then, the problem of finding a minimum-cost balanced bisection of the MRF that results from $\Gamma$ is NP-hard in the size of the evidence (data).*

The proof (Appendix B.6) is by reduction to the graph minimum bisection problem [14], which is hard to approximate (unless P = NP, there is no PTAS). In fact, the problem we are facing (*multi-way hypergraph partitioning*) is more challenging than graph bisection, and has been extensively studied [12, 24]. And so, we design a simple, greedy partitioning algorithm: it assigns each clause to a bin in descending order by clause weight, subject to the constraint that no component in the resulting graph is larger than an input parameter $\beta$. We include pseudocode in Appendix B.7.

*Partition-aware Search.* We need to refine the search procedure to be aware of partitions: the central challenge is that a clause in the cut may depend on atoms in two distinct partitions. Hence, there are dependencies between the partitions. We exploit the idea in Example 2 to design the following partition-aware search scheme – which is an instance of the Gauss-Seidel method from nonlinear optimization [3, pg. 219]. Denote by $X_1, \ldots, X_k$ the states (i.e., truth assignments to the atoms) of the partitions. First initialize $X_i = x_i^0$ for $i = 1 \ldots k$. For $t = 1 \ldots T$, for $i = 1 \ldots k$, run WalkSAT on $x_i^{t-1}$ conditioned on $\{x_j^t | 1 \leq j < i\} \cup \{x_j^{t-1} | i < j \leq k\}$ to obtain $x_i^t$. Finally, return $\{x_i^T | 1 \leq i \leq k\}$.

*Tradeoffs.* Although fine-grained partitioning improves per-partition search speed (Theorem 3.1) and space efficiency, it also increases cut sizes – especially for dense graphs – which would in turn slow down the Gauss-Seidel inference scheme. Thus, there is an interesting tradeoff of partitioning granularity. In Section B.8, we describe a basic heuristic that combines Theorem 3.1 and the Gauss-Seidel scheme.

## 4. EXPERIMENTS

In this section, we validate first that our system TUFFY is orders of magnitude more scalable and efficient than prior approaches. We then validate that each of our techniques contributes to the goal.

*Experimental Setup.* We select ALCHEMY, the currently most widely used MLN system, as our comparison point. ALCHEMY and TUFFY are implemented in C++ and Java, respectively. The RDBMS used by TUFFY is PostgreSQL 8.4. Unless specified otherwise, all experiments are run on an Intel Core2 at 2.4GHz with 4 GB of RAM running Red Hat Enterprise Linux 5. For fair comparison, in all experiments TUFFY runs a single thread unless otherwise noted.

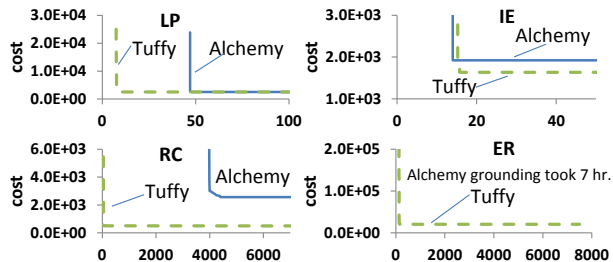

Figure 3: Time-cost plots of Alchemy vs. Tuffy; the $x$ axes are time (sec)

*Datasets.* We run ALCHEMY and TUFFY on four datasets; three of them (including their MLNs) are taken directly from the ALCHEMY website [7]: *Link Prediction* (**LP**), given an administrative database of a CS department, the goal is to predict student-adviser relationships; *Information Extraction* (**IE**), given a set of Citeseer citations, the goal is to extract from them structured records; and *Entity Resolution* (**ER**), which is to deduplicate citation records based on word similarity. These tasks have been extensively used in prior work. The last task, *Relational Classification* (**RC**), performs classification on the Cora dataset [15]; **RC** contains all the rules in Figure 1. Table 1 contains statistics about the data.

|                  | **LP** | **IE** | **RC** | **ER** |
|------------------|--------|--------|--------|--------|
| #relations       | 22     | 18     | 4      | 10     |
| #rules           | 94     | 1K     | 15     | 3.8K   |
| #entities        | 302    | 2.6K   | 51K    | 510    |
| #evidence tuples | 731    | 0.25M  | 0.43M  | 676    |
| #query atoms     | 4.6K   | 0.34M  | 10K    | 16K    |
| #components      | 1      | 5341   | 489    | 1      |

Table 1: Dataset statistics

### 4.1 High-level Performance

We empirically demonstrate that TUFFY with all the techniques we have described has faster grounding, higher search speed, lower memory usage, and in some cases produces much better solutions than a competitor main memory approach, ALCHEMY. Recall that the name of the game is to produce low-cost solutions quickly. With this in mind, we run TUFFY and ALCHEMY on each dataset for 7500 seconds, and track the cost of the best solution found up to any moment; on datasets that have multiple components, namely IE and RC, we apply the partitioning strategy in Section 3.3 on TUFFY. As shown in Figure 3, TUFFY often reaches a best solution within orders of magnitude less time than ALCHEMY; secondly, the result quality of TUFFY is at least as good as – sometimes substantially better (e.g., on IE and RC) than – ALCHEMY. Here, we have zoomed the time axes into interesting areas. Since "solution cost" is undefined during grounding, each curve begins only when grounding is completed[6]. We analyze the experiment results in more detail in the following sections.

---
[6]The L-shaped curves indicate that search converges very quickly compared to grounding time.

|         | LP | IE | RC    | ER     |
|---------|----|----|-------|--------|
| ALCHEMY | 48 | 13 | 3,913 | 23,891 |
| TUFFY   | 6  | 13 | 40    | 106    |

Table 2: Grounding time (sec)

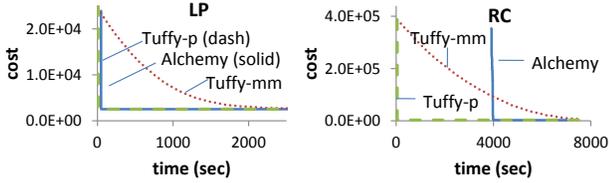

Figure 4: Time-cost plots of Alchemy vs. Tuffy-p (i.e., Tuffy without partitioning) vs. Tuffy-mm (i.e., Tuffy with RDBMS-based search)

## 4.2 Effect of Bottom-up Grounding

We validate that the RDBMS-based grounding approach in TUFFY allows us to complete the grounding process orders of magnitude more efficiently than ALCHEMY. To make this point, we run TUFFY and ALCHEMY on the four datasets, and show their grounding time in Table 2. We can see that TUFFY outperforms ALCHEMY by orders of magnitude at run time in the grounding phase (a factor of 225 on the ER dataset). To understand the differences, we dug deeper with a lesion study (i.e., disabling one aspect of a system at a time), and found that sort join and hash join algorithms (along with predicate pushdown) are the key components of the RDBMS that speeds up the grounding process of TUFFY (Appendix C.2). TUFFY obviates the need for ALCHEMY to reimplement the optimization techniques in an RDBMS.

## 4.3 Effect of Hybrid Architecture

We validate two technical claims: (1) the hybrid memory management strategy of TUFFY (even without our partitioning optimizations) has comparable search rates to existing main memory implementations (and much faster than RDBMS-based implementation) and (2) TUFFY maintains a much smaller memory footprint (again without partitioning). Thus, we compare three approaches: (1) TUFFY without the partitioning optimizations, called TUFFY-p (read: Tuffy minus p), (2) a version of TUFFY (also without partitioning) that implements RDBMS-based WalkSAT (detailed in Appendix B.2), TUFFY-mm, and (3) ALCHEMY.

Figure 4 illustrates the time-cost plots on LP and RC of all three approaches. We see from RC that TUFFY-p is able to ground much more quickly than ALCHEMY (40 sec compared to 3913 sec). Additionally, we see that, compared to TUFFY-mm, TUFFY-p's in-memory search is orders of magnitude faster at getting to their best reported solution (both approaches finish grounding at the same time, and so start search at the same time). To understand why, we measure the *flipping rate*, which is the number of steps performed by WalkSAT per second. As shown in Table 3, the reason is that TUFFY-mm has a dramatically lower flipping rate. We discuss the performance bound of any RDBMS-based search implementation in Appendix C.1.

To validate our second claim, that TUFFY-p has a smaller memory footprint, we see in Table 4, that on all datasets, the memory footprint of TUFFY is no more than 5% of ALCHEMY. Drilling down, the reason is that the intermediate state size of ALCHEMY's grounding process may be larger than the size of grounding results. For example, on the RC dataset, ALCHEMY allocated 2.8 GB of RAM only to produce 4.8 MB of ground clauses. While ALCHEMY has to hold everything in memory, TUFFY only needs to load the grounding result from the RDBMS at the end of grounding. It follows that, given the same resources, there are MLNs that TUFFY can handle efficiently while ALCHEMY would crash. Indeed, on a dataset called "ER+" which is twice as large as ER, ALCHEMY exhausts all 4GB of RAM and crashes soon after launching, whereas TUFFY runs normally with peak RAM usage of roughly 2GB.

|          | LP    | IE    | RC    | ER   |
|----------|-------|-------|-------|------|
| ALCHEMY  | 0.20M | 1M    | 1.9K  | 0.9K |
| TUFFY-mm | 0.9   | 13    | 0.9   | 0.03 |
| TUFFY-p  | 0.11M | 0.39M | 0.17M | 7.9K |

Table 3: Flipping rates (#flips/sec)

|              | LP     | IE     | RC     | ER     |
|--------------|--------|--------|--------|--------|
| clause table | 5.2 MB | 0.6 MB | 4.8 MB | 164 MB |
| ALCHEMY RAM  | 411 MB | 206 MB | 2.8 GB | 3.5 GB |
| TUFFY-p RAM  | 9 MB   | 8 MB   | 19 MB  | 184 MB |

Table 4: Space efficiency of Alchemy vs. Tuffy-p (without partitioning)

From these experiments, we conclude that the hybrid architecture is crucial to TUFFY's overall efficiency.

## 4.4 Effect of Partitioning

In this section, we validate that, when there are multiple components in the data, partitioning not only improves TUFFY's space efficiency, but – due to Theorem 3.1 – may actually enable TUFFY to find substantially higher quality results. We compare TUFFY's performance (with partitioning enabled) against TUFFY-p: a version of TUFFY with partitioning disabled.

We run the search phase on each of the four datasets using three approaches: ALCHEMY, TUFFY-p, and TUFFY (with partitioning). TUFFY-p and ALCHEMY run WalkSAT on the whole MRF for $10^7$ steps. TUFFY runs WalkSAT on each component in the MRF independently, each component $G_i$ receiving $10^7 |G_i|/|G|$ steps, where $|G_i|$ and $|G|$ are the numbers of atoms in this component and the MRF, respectively. This is weighted round-robin scheduling.

|               | LP   | IE   | RC    | ER     |
|---------------|------|------|-------|--------|
| #components   | 1    | 5341 | 489   | 1      |
| TUFFY-p RAM   | 9MB  | 8MB  | 19MB  | 184MB  |
| TUFFY RAM     | 9MB  | 8MB  | 15MB  | 184MB  |
| TUFFY-p cost  | 2534 | 1933 | 1943  | 18717  |
| TUFFY cost    | 2534 | 1635 | 1281  | 18717  |

Table 5: Performance of Tuffy vs. Tuffy-p (i.e., Tuffy without partitioning)

As shown in Table 5, when there are multiple components in the MRF, partitioning allows TUFFY to use less memory than TUFFY-p. (The IE dataset is too small to yield notable differences). We see that TUFFY's component-aware inference can produce significantly better results than TUFFY-p. We then extend the run time of all systems. As shown in Figure 5, there continues to be a gap between TUFFY's component-aware search approach and the original WalkSAT running on the whole MRF. This gap is predicted by our theoretical analysis in Section 3.3. Thus, we have verified that partitioning makes Tuffy substantially more efficient in terms of both space and search speed.

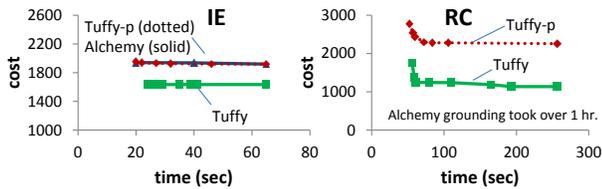

Figure 5: Time-cost plots of Tuffy vs Tuffy-p (i.e., Tuffy without partitioning)

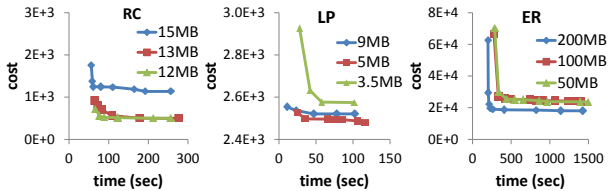

Figure 6: Time-cost plots of Tuffy with different memory budgets

We also validate that Tuffy's loading and parallelism makes a substantial difference: without our batch loading technique, Tuffy takes 448s to perform $10^6$ search steps per component on RC, while 117s to perform the same operation with batch loading. With the addition of 8 threads (on 8 cores), we further reduce the runtime to 28s. Additional loading and parallelism experiments in Appendix C.3 support our claim that our loading algorithm and partitioning algorithm contribute to improving processing speed.

### 4.5 Effect of Further Partitioning

To validate our claim that splitting MRF components can further improve both space efficiency and sometimes also search quality (Section 3.4), we run Tuffy on RC, ER, and LP with different memory budgets – which are fed to the partitioning algorithm as the bound of partition size. On each dataset, we give Tuffy three memory budgets, with the largest one corresponding to the case when no components are split. Figure 6 shows the experiment results. On RC, we see another improvement of the result quality (cf. Figure 5). Similar to Example 2, we believe the reason to be graph sparsity: "13MB" cuts only about 420 out of the total 10K clauses. In contrast, while MRF partitioning lowers RAM usage considerably on ER, it also leads to slower convergence – which correlates with poor partitioning quality: the MRF of ER is quite dense and even 2-way partitioning ("100MB") would cut over 1.4M out of the total 2M clauses. The dataset LP illustrates the interesting tradeoff where a coarse partition is beneficial whereas finer grained partitions would be detrimental. We discuss this tradeoff in Appendix B.8.

### 5. CONCLUSION

Motivated by a large set of data-rich applications, we study how to push MLN inference inside an RDBMS. We find that the grounding phase of MLN inference performs many relational operations and that these operations are a substantial bottleneck in state-of-the-art MLN implementations such as Alchemy. Using an RDBMS, Tuffy not only achieves scalability, but also speeds up the grounding phase by orders of magnitude. We then develop a hybrid solution with RDBMS-based grounding and in-memory search. To improve the space and time efficiency of Tuffy, we study a partitioning approach that allows for in-memory search even when the dataset does not fit in memory. We showed that further partitioning allows Tuffy to produce higher quality results in a shorter amount of time.

# APPENDIX

## A. MATERIAL FOR PRELIMINARIES

### A.1 More Details on the MLN Program

Rules in MLNs are expressive and may involve data in non-trivial ways. For example, consider $F_2$:

$$\texttt{wrote}(x,p1), \texttt{wrote}(x,p2), \texttt{cat}(p1,c) => \texttt{cat}(p2,c) \quad (F_2)$$

Intuitively, this rule says that all the papers written by a particular person are likely to be in the same category. Rules may also have existential quantifiers: $F_4$ in Figure 1 says *"any paper in our database must have at least one author."* It is also a hard rule, which is indicated by the infinite weight, and so no possible world may violate this rule. The weight of a formula may also be negative, which effectively means that the negation of the formula is likely to hold. For example, $F_5$ models our belief that none or very few of the unlabeled papers belong to 'Networking'. TUFFY supports all of these features. If the input MLN contains hard rules (indicated by a weight of $+\infty$ or $-\infty$), then we insist that the set of possible worlds (Inst) only contain worlds that satisfy every hard rule with $+\infty$ and violate every rule with $-\infty$.

### A.2 Markov Random Field

A *Boolean Markov Random Field* (or *Boolean Markov network*) is a model of the joint distribution of a set of Boolean random variables $\bar{X} = (X_1, \ldots, X_N)$. It is defined by a hypergraph $G = (X, E)$; for each hyperedge $e \in E$ there is a *potential function* (aka "feature") denoted $\phi_e$, which is a function from the values of the set of variables in $e$ to non-negative real numbers. This defines a joint distribution $\Pr(\bar{X} = \bar{x})$ as follows:

$$\Pr(\bar{X} = \bar{x}) = \frac{1}{Z} \prod_{e \in E} \phi_e(\bar{x}_e)$$

where $\bar{x} \in \{0,1\}^N$, $Z$ is a normalization constant and $\bar{x}_e$ denotes the values of the variables in $e$.

Fix a set of constants $C = \{c_1, \ldots, c_M\}$. An MLN defines a Boolean Markov Random Field as follows: for each possible grounding of each predicate (i.e., atom), create a node (and so a Boolean random variable). For example, there will be a node $\texttt{refers}(p1,p2)$ for each pair of papers $p1, p2$. For each formula $F_i$ we ground it in all possible ways, then we create a hyperedge $e$ that contains the nodes corresponding to all terms in the formula. For example, the key constraint creates hyperedges for each paper and all of its potential categories.

### A.3 Optimizing MLN Grounding Process

Conceptually, we might ground an MLN formula by enumerating all possible assignments to its free variables. However, this is both impractical and unnecessary. For example, if we ground $F_2$ exhaustively this way, the result would contain $|D|^4$ ground clauses. Fortunately, in practice a vast majority of ground clauses are satisfied by evidence regardless of the assignments to unknown truth values; we can safely discard such clauses [40]. Consider the ground clause $g_{\bar{d}}$ of $F_2$ where $\bar{d} =$('Joe', 'P2', 'P3', 'DB'). Suppose that $\texttt{wrote}$('Joe', 'P3') is known to be false, then $g_{\bar{d}}$ will be satisfied no matter how the other atoms are set ($g_{\bar{d}}$ is an implication). Hence, we can ignore $g_{\bar{d}}$ during the search phase.

Pushing this idea further, [39] proposes a method called "lazy inference" which is implemented by ALCHEMY. Specifically, ALCHEMY works under the more aggressive hypothesis that most atoms will be false in the final solution, and in fact throughout the entire execution. To make this idea precise, call a ground clause *active* if it can be violated by flipping zero or more active atoms, where an atom is active if its value *flips* at any point during execution. Observe that in the preceding example the ground clause $g_{\bar{d}}$ is not active. ALCHEMY keeps only *active ground clauses* in memory, which can be much smaller than the full set of ground clauses. Furthermore, as on-the-fly incremental grounding is more expensive than batch grounding, ALCHEMY uses the following one-step look-ahead strategy: assume all atoms are inactive and compute active clauses; *activate* the atoms in the grounding result and recompute active clauses. This "look-ahead" procedure could be repeatedly applied until convergence, resulting in an *active closure*. TUFFY implements this closure algorithm.

### A.4 The WalkSAT Algorithm

We list the pseudocode of WalkSAT [13] in Algorithm 1.

---

**Algorithm 1** The WALKSAT Algorithm

---
**Input:** $A$: an set of atoms
**Input:** $C$: an set of weighted ground clauses
**Input:** MaxFlips, MaxTries
**Output:** $\sigma^*$: a truth assignment to $A$
1: lowCost $\leftarrow +\infty$
2: **for** try $= 1$ to MaxTries **do**
3:    $\sigma \leftarrow$ a random truth assignment to $A$
4:    **for** flip $= 1$ to MaxFlips **do**
5:      pick a random $\texttt{c} \in C$ that is violated
6:      rand $\leftarrow$ a random float between 0.0 and 1.0
7:      **if** rand $\leq 0.5$ **then**
8:         flip a random atom in $\texttt{c}$
9:      **else**
10:        flip an atom in $\texttt{c}$ s.t. the cost decreases most
11:      **if** cost $<$ lowCost **then**
12:         lowCost $\leftarrow$ cost, $\sigma^* \leftarrow \sigma$

---

### A.5 Marginal Inference of MLNs

In marginal inference, we estimate the marginal probability of atoms. Since this problem is generally intractable, we usually resort to sampling methods. The state-of-the-art marginal inference algorithm is MC-SAT [38], which is implemented in both ALCHEMY and TUFFY. In MC-SAT, each sampling step consists of a call to a heuristic SAT sampler named SampleSAT [44]. Essentially, SampleSAT is a combination of simulated annealing and WalkSAT. And so, TUFFY is able to perform marginal inference more efficiently as well. ALCHEMY also implements a lifted algorithm for marginal inference [42]; it is future work to extend our study to lifted approaches.

## B. MATERIAL FOR SYSTEMS

### B.1 A Compilation Algorithm for Grounding

Algorithm 2 is a basic algorithm of expressing the grounding process of an MLN formula in SQL. To support existential quantifiers, we used PostgreSQL's array aggregate fea-

381

ture. The ideas in Appendix A.3 can be easily implemented on top of this algorithm.

---

**Algorithm 2** MLN Grounding in SQL

---

**Input:** an MLN formula $\phi = \vee_{i=1}^{k} l_i$ where each $l_i$ is a literal supported by predicate table $r(l_i)$
**Output:** a SQL query $Q$ that grounds $\phi$
1: FROM clause of $Q$ includes '$r(l_i)\ t_i$' for each literal $l_i$
2: SELECT clause of $Q$ contains '$t_i.aid$' for each literal $l_i$
3: For each positive (resp. negative) literal $l_i$, there is a WHERE predicate '$t_i.truth \neq$ true' (resp. '$t_i.truth \neq$ false')
4: For each variable $x$ in $\phi$, there is a WHERE predicate that equates the corresponding columns of $t_i$'s with $l_i$ containing $x$
5: For each constant argument of $l_i$, there is an equal-constant WHERE predicate for table $t_i$
6: Form a conjunction with the above WHERE predicates

## B.2 Implementing WalkSAT in RDBMS

WalkSAT is a stochastic local search algorithm; its random access patterns pose considerable challenges to the design of TUFFY. More specifically, the following operations are difficult to implement efficiently with on-disk data: 1) uniformly sample an unsatisfied clause; 2) random access (read/write) to per-atom or per-clause data structures; and 3) traverse clauses involving a given atom. Atoms are cached as in-memory arrays, while the per-clause data structures are read-only. Each step of WalkSAT involves a scan over the clauses and many random accesses to the atoms.

Although our design process iterated over numerous combinations of various design choices, we were still unable to reduce the gap as reported in Section 4.2. For example, compared to clause table scans, one might suspect that indexing could improve search speed by reading less data at each step. However, we actually found that the cost of maintaining indices often outweighs the benefit provided by indexing. Moreover, we found it very difficult to get around RDBMS overhead such as PostgreSQL's mandatory MVCC.

## B.3 Illustrating Tuffy's Hybrid Architecture

Figure 7 illustrates the hybrid memory management approach of TUFFY. ALCHEMY is a representative of prior art MLN systems, which uses RAM for both grounding and search; TUFFY-mm is a version of TUFFY we developed that uses an RDBMS for all memory management; and TUFFY is the hybrid approach as discussed in Section 3.2.

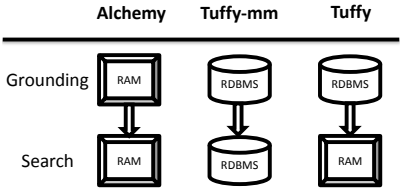

Figure 7: Comparison of architectures

## B.4 MLNs Causing MRF Fragmentation

MLN rules usually model the interaction of relationships and attributes of some underlying entities. As such, one can define entity-based transitive closures, which directly corresponds to components in the MRF. Since in real world data the interactions are usually sparse, one can expect to see multiple components in the MRF. A concrete example is the paper classification running example, where the primary entities are papers, and the interactions are defined by citations and common authors. Indeed, our RC dataset yields hundreds of components in the MRF (see Table 5).

## B.5 Theorem 3.1

PROOF OF THEOREM 3.1. We follow the notations of the theorem. Without loss of generality and for ease of notation, suppose $H = \{1, \ldots, N\}$. Denote by $\Omega$ the state space of $G$. Let $Q_k \subseteq \Omega$ be the set of states of $G$ where there are exactly $k$ non-optimal components. For any state $x \in \Omega$, define $H(x) = \mathbf{E}[H_x(Q_0)]$, i.e., the expected hitting time of an optimal state from $x$ when running WalkSAT. Define $f_k = \min_{x \in Q_k} H(x)$; in particular, $f_0 = 0$, and $f_1$ corresponds to some state that differs from an optimal by only one bit. Define $g_k = f_{k+1} - f_k$. For any $x, y \in \Omega$, let $\Pr(x \to y)$ be the transition probability of WalkSAT, i.e., the probability that next state will be $y$ given current state $x$. Note that $\Pr(x \to y) > 0$ only if $y \in N(x)$, where $N(x)$ is the set of states that differ from $x$ by at most one bit. For any $A \subseteq \Omega$, define $\Pr(x \to A) = \sum_{y \in A} \Pr(x \to y)$.

For any $x \in Q_k$, we have

$$\begin{aligned} H(x) &= 1 + \sum_{y \in \Omega} \Pr(x \to y) H(y) \\ &= 1 + \sum_{t \in \{-1,0,1\}} \sum_{y \in Q_{k+t}} \Pr(x \to y) H(y) \\ &\geq 1 + \sum_{t \in \{-1,0,1\}} \sum_{y \in Q_{k+t}} \Pr(x \to y) f_{k+t}. \end{aligned}$$

Define

$$P_+^x = \Pr(x \to Q_{k+1}), \quad P_-^x = \Pr(x \to Q_{k-1}),$$

then $\Pr(x \to Q_k) = 1 - P_+^x - P_-^x$, and

$$H(x) \geq 1 + f_k(1 - P_+^x - P_-^x) + f_{k-1}P_-^x + f_{k+1}P_+^x.$$

Since this inequality holds for any $x \in Q_k$, we can fix it to be some $x^* \in Q_k$ s.t. $H(x^*) = f_k$. Then $g_{k-1}P_-^{x^*} \geq 1 + g_k P_+^{x^*}$, which implies $g_{k-1} \geq g_k P_+^{x^*}/P_-^{x^*}$.

Now without loss of generality assume that in $x^*$, $G_1, \ldots, G_k$ are non-optimal while $G_{k+1}, \ldots, G_N$ are optimal. Let $x_i^*$ be the projection of $x^*$ on $G_i$. Then since

$$P_-^{x^*} = \frac{\sum_1^k v_i(x_i^*)\alpha_i(x_i^*)}{\sum_1^N v_i(x_i^*)}, \quad P_+^{x^*} = \frac{\sum_{k+1}^N v_j(x_j^*)\beta_j(x_j^*)}{\sum_1^N v_i(x_i^*)},$$

we have

$$g_{k-1} \geq g_k \frac{\sum_{k+1}^N v_j(x_j^*)\beta_j(x_j^*)}{\sum_1^k v_i(x_i^*)\alpha_i(x_i^*)} \geq g_k \frac{r(N-k)}{k},$$

where the second inequality follows from the definition of $r$.

For all $k \leq rN/(r+2)$, we have $g_{k-1} \geq 2g_k$. Since $g_k \geq 1$ for any $k$, $f_1 = g_0 \geq 2^{rN/(r+2)}$. That is, not aware of components, WalkSAT would take an exponential number of steps in expectation to correct the last bit to reach an optimum. □

According to this theorem, the gap on Example 1 is at least $2^{N/3}$; in fact, a more detailed analysis reveals that the gap is at least $\binom{N-1}{\frac{N}{2}} \approx \Theta(2^N/\sqrt{N})$. Figure 8 shows the experiment results of running ALCHEMY, TUFFY, and TUFFY-p (i.e., TUFFY without partitioning) on Example 1 with 1000 components. Note that the analysis of Theorem 3.1 actually applies to not only WalkSAT, but stochastic local search in general. Since stochastic local search algorithms are used in many statistical models, we believe that our observation here and corresponding techniques have much wider implications than MLN inference.

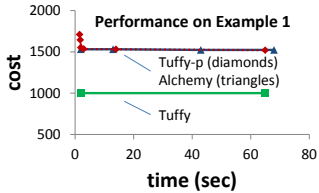

**Figure 8: Effect of partitioning on Example 1**

### B.6 Hardness of MRF Partitioning

A bisection of a graph $G = (V, E)$ with an even number of vertices is a pair of disjoint subsets $V_1, V_2 \subset V$ of equal size. The cost of a bisection is the number of edges adjacent to both $V_1$ and $V_2$. The problem of *Minimum Graph Bisection* (MGB) is to find a bisection with minimum cost. This problem admits no PTAS [14]. The hardness of MGB directly implies the hardness of partitioning MRFs. As such, one may wonder if it still holds w.r.t. the domain size for a *given* MLN program (hence of size $O(1)$). The following theorem shows that the answer is yes.

THEOREM B.1. *MGB can be reduced to the problem of finding a minimum bisection of the MRF generated an MLN of size $O(1)$.*

PROOF. Consider the MLN that contains a single formula of the following form:

$$p(x), r(x, y) \to p(y),$$

where $p$ is query and $r$ is evidence. For any graph $G = (V, E)$, we can set the domain of the predicates to be $V$, and let $r = E$. The MRF generated by the above MLN (using techniques in Appendix A.3) is identical to $G$. □

### B.7 MRF Partitioning Algorithm

We provide a very simple MRF partitioning algorithm (Algorithm 3) that is inspired by Kruskal's minimum spanning tree algorithm. It agglomeratively merges atoms into partitions with one scan of the clauses sorted in the (descending) absolute values of weights. The hope is to avoid cutting high-weighted clauses, thereby (heuristically) minimizing weighted cut size.

To explain the partitioning procedure, we provide the following definitions. Each clause $c$ in the MRF $G = (V, E)$ is *assigned* to an atom in $c$. A partition of the MRF is a subgraph $G_i = (V_i, E_i)$ defined by a subset of atoms $V_i \subseteq V$; $E_i$ is the set of clauses assigned to some atom in $V_i$. The size of $G_i$ as referred to by Algorithm 3 can be any monotone function in $G_i$; in practice, it is defined to be the total number of literals and atoms in $G_i$. Note that when the parameter $\beta$ is set to $+\infty$, the output is the connected components of $G$.

Our implementation of Algorithm 3 only uses RAM to maintain a union-find structure of the nodes, and performs all other operations in the RDBMS. For example, we use SQL queries to "assign" clauses to atoms and to compute the partition of clauses from a partition of atoms.

**Algorithm 3** A Simple MRF Partitioning Algorithm
**Input:** an MRF $G = (V, E)$ with clause weights $w : E \mapsto \mathbb{R}$
**Input:** partition size bound $\beta$
**Output:** a partitioning of $V$ s.t. the size of each partition is no larger than $\beta$
1: Initialize hypergraph $H = (V, F)$ with $F = \emptyset$
2: **for all** $e \in E$ in $|w|$-descending order **do**
3: $\quad F \leftarrow F \cup e$ if afterwards no component in $H$ is larger than $\beta$
4: **return** the collection of per-component atom sets in $H$

### B.8 Tradeoff of MRF Partitioning

Clearly, partitioning might be detrimental to search speed if the cut size is large. Furthermore, given multiple partitioning options, how do we decide which one is better? As a baseline, we provide the following formula to (roughly) estimate the benefit (if positive) or detriment (if negative) of a partitioning:

$$W = 2^{\frac{N}{3}} - T\frac{|\#cut\_clauses|}{|E|},$$

where $N$ is the estimated number of components with positive lowest cost, T is the total number of WalkSAT steps in one round of Gauss-Seidel, and $|E|$ is the total number of clauses. The first term roughly captures the speed-up as a result of Theorem 3.1, and the second term roughly captures the slow-down caused by cut clauses.

Empirically however, we find this formula to be rather conservative compared to experimental results that generally favor much more aggressive partitioning. In the technical report [16] (Section 5), we present a much more detailed discussion. The main idea is to finely model the elements of the tradeoff by taking into account connectivity and the influence of individual atoms.

## C. MATERIAL FOR EXPERIMENTS

### C.1 Alternative Search Algorithms

As shown in Section 4.3, RDBMS-based implementation of WalkSAT is several orders of magnitude slower than the in-memory counter part. This gap is consistent with the I/O performance of disk vs. main memory. One might imagine some clever caching schemes for WalkSAT, but even assuming that a flip incurs only one random I/O operation (which is usually on the order of 10 ms), the flipping rate of RDBMS-based search is still no more than 100 flips/sec. Thus, it is highly unlikely that disk-based search implementations could catch up to their in-memory counterpart.

### C.2 Lesion Study of Tuffy Grounding

To understand which part of the RDBMS contributes the most to TUFFY's fast grounding speed, we conduct a lesion study by comparing the grounding time in three settings: 1) **full optimizer**, where the RDBMS is free to optimize SQL queries in all ways; 2) **fixed join order**, where we force the RDBMS to use the same join order as ALCHEMY does; 3) **fixed join algorithm**, where we force the RDBMS to use nested loop join only. The results are shown in Table 6.

|  | LP | IE | RC | ER |
|---|---|---|---|---|
| Full optimizer | 6 | 13 | 40 | 106 |
| Fixed join order | 7 | 13 | 43 | 111 |
| Fixed join algorithm | 112 | 306 | >36,000 | >16,000 |

Table 6: Grounding time in seconds

|  | IE | RC |
|---|---|---|
| Tuffy-batch | 448 | 133 |
| Tuffy | 117 | 77 |
| Tuffy+parallelism | 28 | 42 |

Table 7: Comparison of execution time in seconds

Clearly, being able to use various join algorithms is the key to TUFFY's fast grounding speed.

### C.3 Data Loading and Parallelism

To validate the importance of batch data loading and parallelism (Section 3.3), we run three versions of TUFFY on the IE and RC datasets: 1) **Tuffy**, which has batch loading but no parallelism; 2) **Tuffy-batch**, which loads components one by one and does not use parallelism; and 3) **Tuffy+parallelism**, which has both batch loading and parallelism. We use the same WalkSAT parameters on each component (up to $10^6$ flips per component) and run all three settings on the same machine with an 8-core Xeon CPU. Table 7 shows the end-to-end running time of each setting.

Clearly, loading the components one by one incurs significant I/O cost on both datasets. The grounding + partitioning time of IE and RC are 11 seconds and 35 seconds, respectively. Hence, Tuffy+parallelism achieved roughly 6-time speed up on both datasets.

### D. EXTENDED RELATED WORK

The idea of using the stochastic local search algorithm WALKSAT to find the most likely world is due to Kautz et al. [13]. Singla and Domingos [41] proposed lazy grounding and applies it to WALKSAT, resulting in an algorithm called LAZYSAT that is implemented in ALCHEMY. The idea of ignoring ground clauses that are satisfied by evidence is highlighted as an effective way of speeding up the MLN grounding process in Shavlik and Natarajan [40], which formulates the grounding process as nested loops and provides heuristics to approximate the optimal looping order. Mihalkova and Mooney [35] also employ a bottom-up approach, but they address structure learning of MLNs whereas we focus on inference. As an orthogonal approach to scaling MLN inference, Mihalkova and Richardson [36] study how to avoid redundant computation by clustering similar query literals. It is an interesting problem to incorporate their techniques into TUFFY. Lifted inference (e.g., [42]) involves performing inference in MLNs without completely grounding them into MRFs. It is interesting future work to extend TUFFY to perform lifted inference. Knowledge-based model construction [28] is a technique that, given a query, finds the minimal relevant portion of a graph; although the resulting subgraph may contain multiple components, the downstream inference algorithm may not be aware of it and thereby cannot benefit from the speedup in Thm. 3.1.

While TUFFY employs the simple WalkSAT algorithm, there are more advanced techniques for MAP inference [31, 33]; we plan to integrate them into upcoming versions of TUFFY. For hypergraph partitioning, there are established solutions such as hMETIS [12]. However, existing implementations of them are limited by memory size, and it is future work to adapt these algorithms to on-disk data; this motivated us to design Algorithm 3. The technique of *cutset conditioning* [17] from the SAT and probabilistic inference literature is closely related to our partitioning technique [30,37]. Cutset conditioning recursively conditions on cutsets of graphical models, and at each step exhaustively enumerates all configurations of the cut, which is impractical in our scenario: even for small datasets, the cut size can easily be thousands, making exhaustive enumeration infeasible. Instead, we use a Gauss-Seidel strategy, which we show is efficient and effective. Additionally, our conceptual goals are different: our goal is to find an analytic formula that quantifies the effect of partitioning and then, we use this formula to optimize the IO and scheduling behavior of a class of local search algorithms; in contrast, prior work focuses on designing new inference algorithms.

There are statistical-logical frameworks similar to MLNs, such as Probabilistic Relational Models [32] and Relational Markov Models [43]. Inference on those models also requires grounding and search, and we are optimistic that the lessons we learned with MLNs carry over to these models.

### E. REFERENCES